\pdfoutput=1
\documentclass[prb,twocolumn,showpacs,amsmath,amssymb,floatfix,superscriptaddress]{revtex4}

\usepackage{graphicx}
\usepackage{dcolumn}
\usepackage{bm}
\usepackage{amsmath}
\usepackage{amssymb,xcolor,soul}
\usepackage{textcomp}
\usepackage[colorlinks=true]{hyperref}
\hypersetup{urlcolor={blue},linkcolor={blue}}

\begin{document}
	
\title{Noncollinear Magnetic Modulation of Weyl Nodes in Ferrimagnetic Mn$_3$Ga}
\author{Cheng-Yi Huang}
\affiliation{Institute of Physics, Academia Sinica, Taipei 11529, Taiwan}
\affiliation{Department of Physics and Astronomy, California State University, Northridge, CA 91330-8268, USA}

\author {Hugo Aramberri}
\email{hugo.aramberri@list.lu}
\affiliation{Department of Physics and Astronomy, California State University, Northridge, CA 91330-8268, USA}

\author{Hsin Lin}
\affiliation{Institute of Physics, Academia Sinica, Taipei 11529, Taiwan}

\author{Nicholas Kioussis}
\email{nick.kioussis@csun.edu }
\affiliation{Department of Physics and Astronomy, California State University, Northridge, CA 91330-8268, USA}
\date{\today}

\begin{abstract}
{The tetragonal ferrimagnetic Mn$_3$Ga exhibits a wide range of intriguing magnetic properties. Here, we report the emergence of topologically nontrivial nodal lines in the absence of spin orbit coupling (SOC) which are protected by both mirror and $C_{4z}$ rotational symmetries.
In the presence of SOC we demonstrate that the doubly degenerate nontrivial crossing points evolve into $C_{4z}$-protected Weyl nodes with chiral charge of $\pm$2. Furthermore, we have considered the experimentally reported noncollinear ferrimagnetic structure, where the magnetic moment of the Mn$_I$ atom (on the Mn-Ga plane) is tilted by an angle $\theta$ with respect to the crystallographic $c$ axis. The evolution of the Weyl nodes with $\theta$ reveals that the double Weyl nodes split into a pair of charge-1 Weyl nodes  whose separation can be tuned by the magnetic orientation in the noncollinear ferrimagnetic structure.}
\end{abstract}	
\maketitle

\section{INTRODUCTION}
The discovery of topological states of matter represents a cornerstone of condensed-matter physics 
that may accelerate the development of quantum information and spintronics and pave the way to realize massless particles such as Dirac and Weyl fermions. A Weyl semimetal (WSM) is a topological semimetallic material hosting doubly-degenerate gapless nodes near the Fermi level in the three-dimensional (3D) momentum space\cite{PTang17,BJYang14,NPArmitage18,MZHasan15}. The nodes correspond to effective magnetic monopoles or antimonopoles which carry nonvanishing positive and negative chiral charge $\pm q$.  Typically, $q$ takes values of $\pm$1 corresponding to Weyl nodes, but is also possible to have integers, $q = \pm2, \pm3, \dots$ for double Weyl nodes, etc. \cite{Hasan16}
The Weyl nodes gives rise to surface states which form open Fermi arcs rather than closed loops. 

Compared to their Dirac semimetal counterparts, WSMs require the breakdown either of inversion symmetry or time reversal symmetry (TRS) to split each four-fold degenerate Dirac node into a pair of Weyl nodes. 
A number of WSMs that break inversion symmetry have been identified in the past few years\cite{PTang17,BJYang14,NPArmitage18,MZHasan15}. 
Moreover the presence of crystalline symmetries can further protect multiple Weyl nodes with large chiral charge\cite{CFang12,ZGao16,BBradlyn17}.
On the other hand, the discovery of their broken TRS  counterparts, 
which link the two worlds of topology and spintronics,  remains challenging and elusive\cite{CFang12}. 
Many potential TRS-breaking WSM have been proposed.  
Recently, three groups have provided unambiguous and direct experimental confirmation that 
Co$_3$Sn$_2$S$_2$\cite{Morali,Belopolskiarxiv}, which becomes a ferromagnet below 175 K, and Co$_2$MnGa, a room-temperature ferromagnet\cite{Belopolski}, are TRS-breaking WSMs. 
The discovery of magnetic WSMs give rise to exotic quantum states ranging from quantum anomalous Hall effect to axion insulators\cite{NPArmitage18}.

Another remarkable and highly promising class of magnetic materials is the Heusler family\cite{CFelser15,JWinterlik12} which  
includes half metals,\cite{RGroot83} ferromagnets, ferrimagnets, antiferromagnets, 
and even topological insulators\cite{SChadov10,HLin10} and Weyl semimetals.  
In particular the ferrimagnetic and antiferromagnetic  compounds with antiparallel exchange coupling, have recently garnered intense interest because of the faster spin dynamics (in the terahertz range) compared to the gigahertz-range magnetization dynamics of their ferromagnetic counterparts.\cite{KCai2019}

The Mn$_3$X (X=Ga, Ge, Sn) Heusler compounds are considered prototypes with promising applications in the area of spintronics\cite{JWinterlik12,DZhang13}. 
These compounds can be experimentally stabilized in either the hexagonal 
DO$_{19}$ structure ($\epsilon$ phase) or the tetragonal DO$_{22}$ structure ($\tau$ phase)\cite{SKhmelevskyi16}. 
The high-temperature hexagonal crystal structure is antiferromagnetic with a high N\'{e}el temperature 
($\sim$ 470 K) and a noncollinear triangular magnetic structure. 
Recently, several experimental and theoretical studies have demonstrated\cite{Kiyohara16,AKNayak16,PKRout19,HYang17,KKuroda17,ZHLiu17,ELiu18} 
the emergence of large anomalous Hall effect (AHE) in the noncollinear AFM hexagonal Mn$_3$X  family, 
whose origin lies on the nonvanishing Berry curvature in momentum space. 
In addition, {\it ab initio} calculations have revealed that these chiral AFM materials are 
topological Weyl semimetals\cite{HYang17}. 
On the other hand, the low-temperature tetragonal phase, 
which can be obtained by annealing the 
hexagonal phase, is ferrimagnetic at room temperature and shows a unique combination of magnetic
and electronic properties, including low magnetization,\cite{Wu2009} 
high uniaxial anisotropy,\cite{Bai2012} high spin polarization ($\approx$ 88\%),\cite{Mizukami2011,Kurt2011,Winterlik2008}
low Gilbert damping constant,\cite{Mizukami2011} high Curie temperature,\cite{Kren1970}  
and large voltage controlled magnetic anisotropy efficiency\cite{Sun2020}.
Interestingly, neutron scattering experiments have reported\cite{KRode13} a noncollinear ferrimagnetic magnetic structure in Mn$_3$Ga, where the magnetic moment orientation of the Mn atoms on the Mn-Ga (001) plane is tilted by about 
21\textdegree with respect to the crystallographic $c$ axis.   


The objective of this work is to carry out first-principles electronic structure calculations to investigate the emergence of topological nodal lines in the absence or presence of SOC in tetragonal ferrimagnetic Mn$_3$Ga. Furthermore, we present results of the effect of non-collinear magnetism on the evolution of the Weyl nodes.




\section{METHODOLOGY}
The electronic structure calculations were carried out by means of first-principles spin-polarized collinear calculations within the density functional theory (DFT) framework as implemented in the VASP package~\cite{Vasp}. The Perdew-Burke-Ernzerhof~\cite{PBEgga} (PBE) implementation of the generalized gradient approximation (GGA) for the exchange-correlation functional was employed. The plane-wave cutoff energy was set to 400 eV, which was enough to yield well-converged results. The Brillouin zone (BZ) was sampled using a $\Gamma$-centered mesh of $10$x$10$x$10$ k-points. 
 The structure was allowed to relax until residual atomic forces became lower than 0.01 eV/\AA\ and residual stresses became smaller than 0.01 GPa. The electron-electron interactions are included, where indicated, within the GGA+U approach of Dudarev et al.~\cite{dudarev1998electron}. In this way, the electron correlations are taken into account through a single effecive parameter $U_\mathrm{eff} = U - J$. The values of $U_\mathrm{eff}$ for the Mn$_{I}$, Mn$_{II}$ and Ga are set to 2.6 eV, 0 and 0, respectively. Previous studies have shown that these values yield lattice parameters closer to their experimental values~\cite{saha2017impact}.
 The spin-orbit coupling (SOC) of the valence electrons is in turn included self-consistently using the second-variation method employing the scalar-relativistic eigenfunctions of the valence states\cite{koelling}, as implemented in VASP.  
Then, DFT derived wave functions both without and with SOC were in turn projected to Wannier functions using the wannier90 package
~\cite{Wannier90}.

In the $D$O$_{22}$ structure (I4/mmm space group) the two (001) antiferromagnetically-coupled Mn sublattices, shown in Fig.~\ref{struc}(a), consist of Mn$_I$ atoms at the Wyckoff positions 2b (0,0,1/2) [Mn$_I$-Ga (001) plane] and Mn$_{II}$ atoms at the 4d (0,1/2,1/4) positions [Mn$_{II}$-Mn$_{II}$ (001) plane]. For the noncollinear calculation, where the magnetic moment of the Mn$_I$ is rotated by an angle $\pi-\theta$ with respect to the [001] direction, 
the angular dependence of the Wannier Hamiltonian in the presence of SOC is determined from,
\begin{equation}
H({\bf k},\theta)=H_0({\bf k})+U(\theta)H_{ex}({\bf k})U^{\dag}(\theta).
\end{equation}
\noindent
Here, $H_0({\bf k})$ is the TRS preserving Hamiltonian with SOC,  
[$TH_0({\bf k})T^{-1}=H_0(-{\bf k})$],
$H_{ex}({\bf k})$ is the TRS-breaking exchange Hamiltonian,
[$TH_{ex}({\bf k})T^{-1}=-H_{ex}(-{\bf k})$], $T=i\sigma_2K$ is the TRS operator, $\sigma_2$ acts on the spin degrees of freedom, $K$ is complex conjugation,
$U(\theta)=e^{-i\frac{\pi-\theta}{2}\sigma_{2,Mn_{I}}}$ is the spin 
rotation operator, and $\sigma_{2,{\text{Mn$_I$}}}$ is the $y$ component of Pauli matrix acting on the spin degrees of freedom of Mn$_I$.

\section{results and discussion}
\subsection{\bf Nodal lines in the absence of SOC}
The calculated lattice parameters $a=b$ = 3.78 \AA\ and $c$ = 7.08 \AA, are in good agreement with previous calculations~\cite{Balke2007,KRode13,Aqtash2015}, which are, however, lower than the experimental values of $a=b$ = 3.92 \AA\ and $c$ = 7.08 \AA (see Table~\ref{Tlat}). 
The effect of U on the topology of the band structure is discussed in Sec. III.
Overall, our calculated GGA values of the magnetic moments of -2.83 $\mu_B$  and 2.30 $\mu_B$ for the Mn$_I$ and Mn$_{II}$ atoms, respectively, are in good agreement with previous DFT calculations\cite{Balke2007,KRode13,Aqtash2015}.
\begin{table}
\caption{List of {\it ab initio} and experimental lattice constants and magnetic moments values for the collinear case. $\mu_z^{2b}$ ($\mu_z^{4d}$) denotes the $z$-component of the magnetic moment of the Mn$_I$ (Mn$_{II}$) atom.}\label{Tlat}
\begin{ruledtabular}
	\begin{tabular}{ccccc}
		Method&a(\AA)&c(\AA)&$\mu_z^{2b}$($\mu_B$/Mn)&$\mu_z^{4d}$ ($\mu_B$/Mn)\\[2pt]
		\hline
		GGA  &3.78&7.08&-2.83&2.30\\
		GGA+U&3.91&7.00&-3.76&2.45\\
		GGA+U+SOC&3.91&7.00&-3.87&2.51\\
		experiment\cite{KRode13}&3.92&7.08&-3.07&2.08
	\end{tabular}	
\end{ruledtabular}
\end{table}

\begin{figure}[t]
	\includegraphics[width=8.cm]{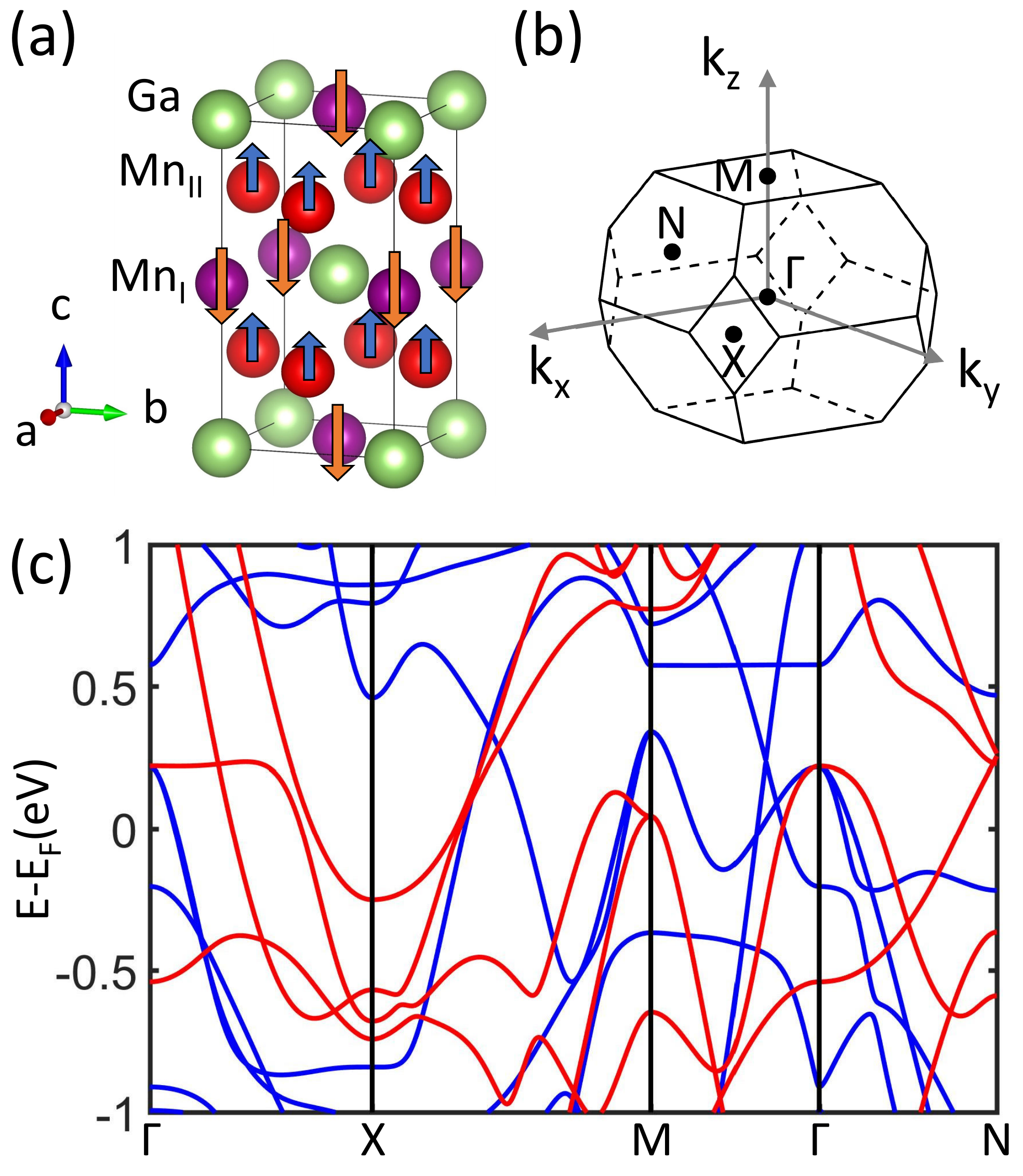} 
	\caption{(Color online) (a) The tetragonal cell of the DO$_{22}$ ferrimagnetic structure with [001] spin polarization. Arrows denote the magnetic moments of Mn$_I$ (purple) and Mn$_{II}$ (red) sublattices which are  coupled antiferromagnetically.  (b) First Brillouin zone of the primitive cell shown in panel (a). (c) Spin-polarized band structure without SOC along the high symmetry directions of the primitive cell, where the spin-up (spin-down) bands are denoted by blue (red).  
		\label{struc}}
\end{figure}
\begin{figure}[h]
	\includegraphics[width=8.cm]{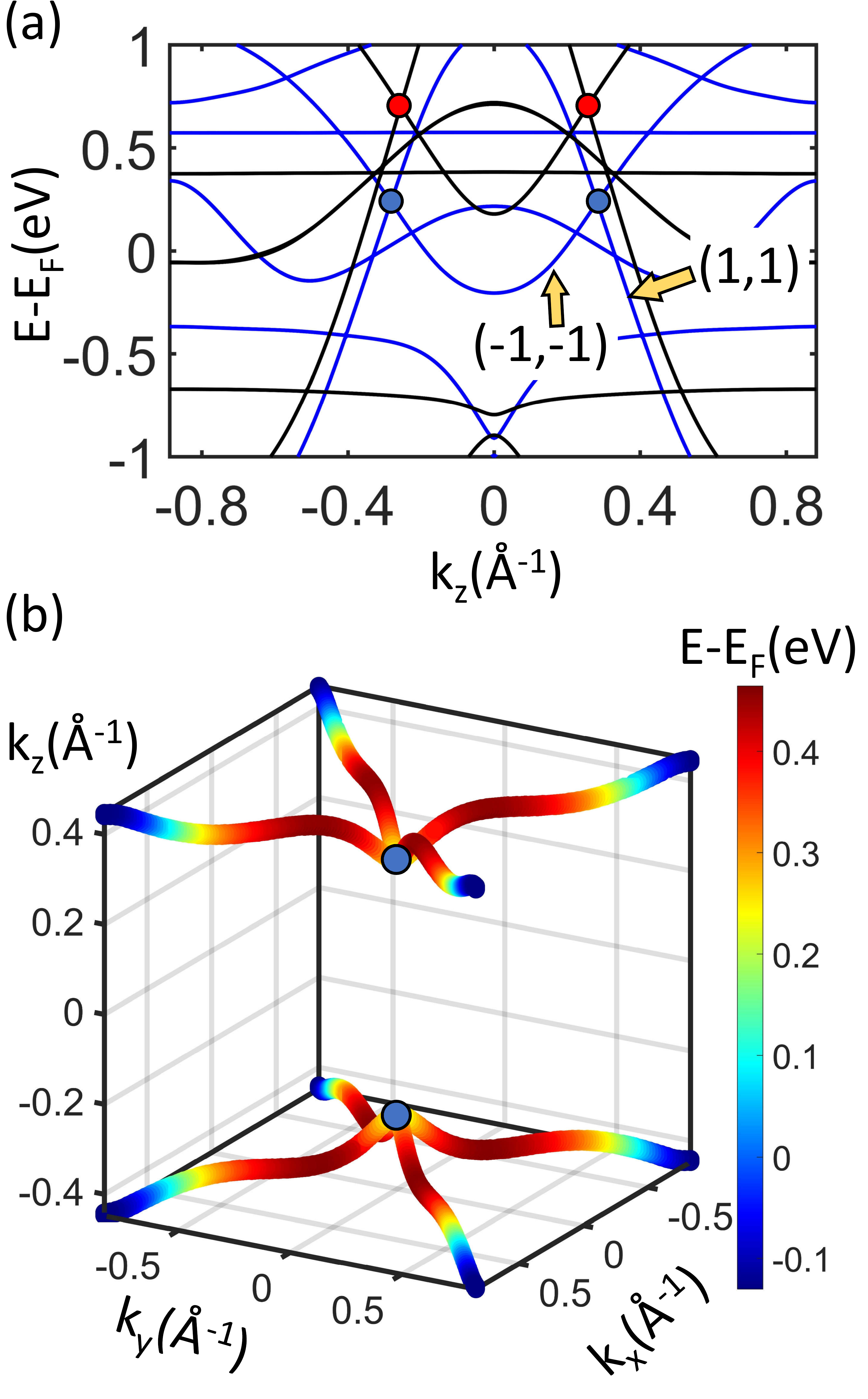} 
	\caption{(Color online) (a) Spin polarized band structure along the $k_z$-axis ($\Gamma-M$ symmetry direction) without SOC, where the blue (black) bands denote the spin-up states calculated from GGA (GGA+U). The two nontrivial crossing points, denoted with the blue dots, are labeled with the pair of eigenvalues, ($\pm1$,$\pm$1), of the mirror, $M_{[110]}$, and four-fold rotational, $C_{4z}$, symmetries, respectively, which protect them. Red dots denote the nontrivial crossing points when U is turned on.
		(b) 3D landscape of the nodal lines where the two blue dots denote the two nontrivial crossing points in (a). The color bar represents the energy of the nodal points relative to the Fermi energy. \label{NL}}
\end{figure}
 Fig.~\ref{struc}(c) shows the spin-polarized band structure of the majority- (blue) and minority-spin (red) bands of Mn$_3$Ga without SOC and with collinear spins along the symmetry lines of the Brillouin zone (BZ) of the primitive cell, shown in Fig.~\ref{struc}(b). For each spin channel, the energy bands can be labeled by the eigenvalues of the crystalline symmetry operator of a particular high symmetry direction. The band structure along the M-$\Gamma$-M direction, shown in Fig.~\ref{NL}(a), features several band crossings close to the Fermi level. Thus, throughout the remainder of the manuscript,  we only focus on the crossing points, marked by blue dots in Fig.~\ref{NL} (a),  between the majority-spin bands along the $k_z$ ($\Gamma-M$) direction. These points are protected by 
 both a mirror reflection symmetry normal to the [110] direction, $M_{[110]}$,  and a four-fold rotational symmetry, $C_{4z}$, and hence can be labeled by the pair of eigenvalues, ($\pm 1$,$\pm 1$), of $M_{[110]}$ and $C_{4z}$, respectively.
 The effective $k\cdot p$ model in the basis $\{|(1,1)\rangle,|(-1,-1)\rangle\}$ up to order of $k^2$ in the absence of SOC can be straightforward derived and is given by
 \begin{equation}
 H_{\text{NL}}=(m_1-m_2k_z^2)s_3+a(k_x^2-k_y^2)s_1,\label{NLeq}
 \end{equation}
where $\bm{k}$ is close to the $\Gamma$ point, $m_1m_2>0$, the $s_i$'s are Pauli matrices and the nodal lines lie on $k_z=\pm\sqrt{\frac{m_1}{m_2}}$ and $k_x=\pm k_y$.
We have tracked the nodal lines on the $M_{[110]}$-invariant plane. The other nodal lines on the $M_{[1\bar{1}0]}$-invariant plane were determined using the $C_{4z}$ rotational symmetry.  
 Fig.~\ref{NL}(b) shows the 3D landscape of nodal lines in momentum space. We find that the nodal lines are topologically nontrivial characterized by the $\pi$ Berry phase\cite{RYu11,CKChiu14,CFang15}. The two blue points denote the nontrivial crossing points as well as the intersecting points of nodal lines along the $k_z$ direction in Fig.~\ref{NL}(b). 
 Notably the crossing points remain gapless and robust against a distortion breaking either $M_{[110]}$ or $C_{4z}$.

\subsection{Weyl Nodes in the Presence of SOC}
In the presence of SOC, the symmetry conservation depends on the magnetic orientation and the crystalline symmetries. More specifically  
the [001] collinear magnetic configuration is invariant under (1) inversion symmetry ($P$), (2) fourfold rotational symmetry about the 
$z$-axis ($C_{4z}$) and (3) mirror reflection symmetry normal to the $z$ direction ($M_z$). We next discuss the effect of magnetization orientation
(collinear versus noncollinear) on the topological features of the band structure.

\begin{table}
	\caption{\label{T1} The $C_{4z}$-protected Weyl fermion on $k_z$ axis. $u_c$$(u_v$) denotes the eigenvalue of $C_{4z}$ in conduction (valence) band. $C$ denotes chiral charge.}
	\begin{ruledtabular}
		\begin{tabular}{c>{\centering}m{2cm}ccm{2.5cm}}
			$k_z$(\AA$^{-1}$) & E-E$_F$ (meV) & $u_c/u_v$ & $C$ & Dispersion \par on $k_x$-$k_y$ plane \\
			\hline\\[-5pt]
			\hspace{3pt}0.2811 & 229 & -1 & 2 & \hspace{1cm} $k^2$\\
			-0.2811 & 229 & -1 & -2 & \hspace{1cm} $k^2$\\
		\end{tabular}
	\end{ruledtabular}
\end{table}

\textit{Weyl Nodes in Collinear Ferrimagnetism---}
\begin{figure*}[tbh]
	\includegraphics[width=0.9\textwidth]{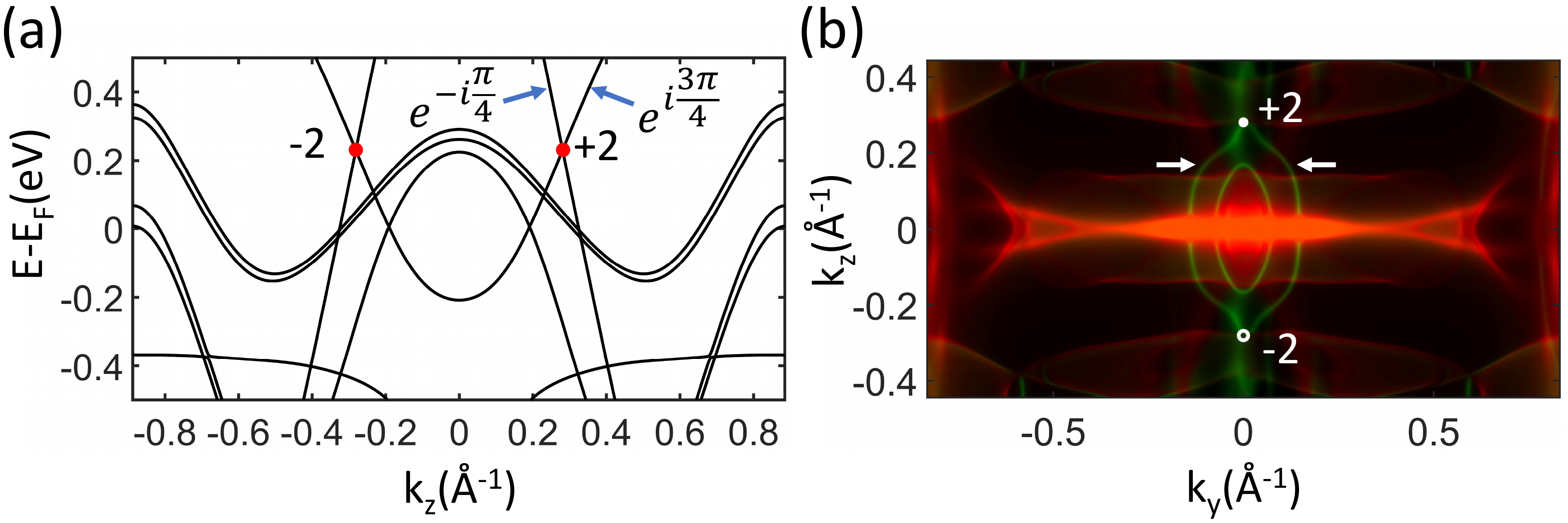} 
	\caption{(Color online) (a) Band structure along the $k_z$-axis ($\Gamma-M$ symmetry direction) with SOC, protected by $C_{4z}$ symmetry. The two Weyl points, denoted with red dots, have chiral charge of $\pm2$. The $e^{-i\frac{\pi}{4}}$ and $e^{i\frac{3\pi}{4}}$ indicate the eigenvalues of $C_{4z}$ for the crossing bands, respectively.
		(b) Two Fermi arcs on the (100) surface where green (red) color denotes the spectral weight of the surface (bulk) states. Solid (hollow) white circle denotes positive (negative) chiral charge, and white arrows indicate the Fermi arcs emerging from the Weyl nodes.}\label{WP}
\end{figure*}
In the presence of SOC, the mirror symmetry $M_{[110]}$ is no longer preserved when the magentization of the collinear ferrimagentic Mn$_3$Ga is along the [001] direction. Consequently,  in general the nodal points in Fig.~\ref{NL}(b) are gapped out except for those crossing points along  
$k_z$ which are protected by the $C_{4z}$ rotational symmetry\cite{com1}. Thus, for the band structure along the $C_{4z}$-invariant $k_z$-axis,
shown in Fig.~\ref{WP}(a), we can identify the states by the eigenvalues of $C_{4z}$ and locate the nontrivial crossing points  associated with different eigenvalues. The nontrivial crossing points, marked by red circles in Fig.~\ref{WP}(a), are Weyl nodes protected by $C_{4z}$ symmetry, whose position along $k_z$, energy relative to E$_F$, ratio of conduction to valence band $C_{4z}$ eigenvalues, $u_c/u_v$, chiral charge, $C$, and dispersion are summarized in Table~\ref{T1}.  Interestingly, the $C_{4z}$-protected Weyl fermion with $u_c/u_v$= -1 carries chiral charge +2 and has quadratic dispersion on the $k_x$-$k_y$ plane,\cite{CFang12} in sharp contrast to the double Weyl fermion with fourfold degeneracy and linear dispersion\cite{PTang17}. Its other parity partner has opposite chiral charge of -2. 
Based on the above analysis, the effective $k\cdot p$ model in the basis of the $C_{4z}$ eigenstates, $\{|e^{-i\frac{\pi}{4}}\rangle,\;|e^{i\frac{3\pi}{4}}\rangle\}$, up to order of $k^2$ for the $C_{4z}$-protected Weyl nodes can be straightforward adapted from Ref.~\onlinecite{CFang12}, and reads
\begin{equation}
H_{\text{WP}}=(m_1-m_2k_z^2)s_3+(ak_+^2+bk_-^2)s_++(ak_-^2+bk_+^2)s_-,
\end{equation}
where $\bm{k}$ is around the $\Gamma$ point, $k_{\pm}=k_x\pm ik_y$, $s_{\pm}=s_1\pm is_2$, $|a|\neq|b|$, $m_1m_2>0$ and the Weyl node positions
are at $k_z=\pm\sqrt{\frac{m_1}{m_2}}$. Interestingly, if $a=b$, $H_{\text{WP}}$ reduces to $H_{\text{NL}}$ in Eq.~(\ref{NLeq}), implying that the gap opening of the nodal lines would close. Fig.~\ref{WP}(b) displays the 
two Fermi arcs on the (100) surface emerging from the two charge-2 Weyl nodes.

\textit{Evolution of Weyl Fermions in NonCollinear Ferrimagnetism---}

\begin{figure}[tbh]
	\includegraphics[width=8.5cm]{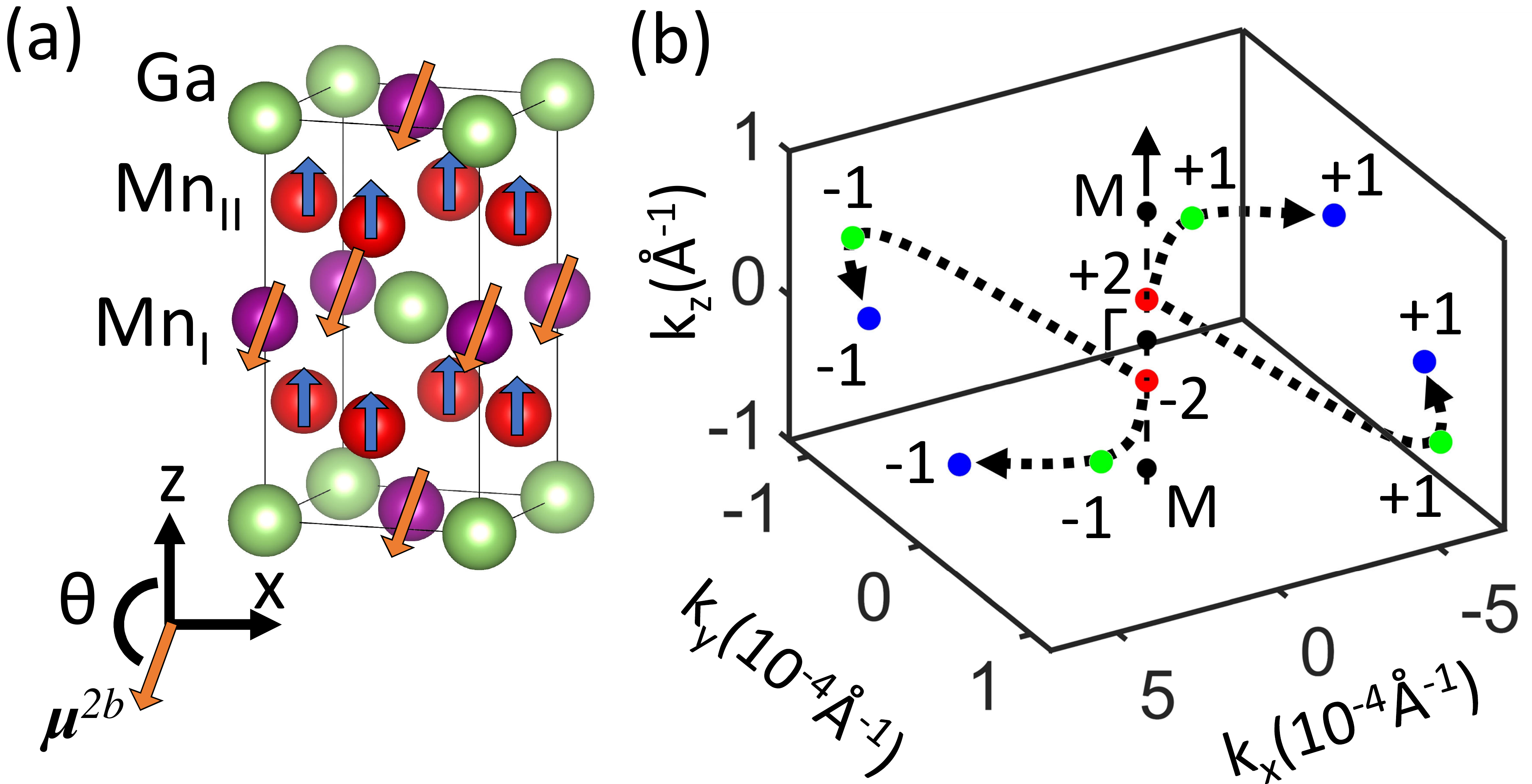} 
	\caption{(Color online) (a) Noncollinear ferrimagnetic DO$_{22}$ structure of Mn$_3$Ga,\cite{KRode13} where the Mn$_I$ atoms [on the Mn-Ga (001) plane] carry a substantial in-plane magnetic moment 
		leading to a tilt of their moments from the crystallographic $c$ axis.
		(b) Evolution of Weyl nodes in the 3D BZ as a function of tilt angle $\theta$, 
		where the red, green and blue circles denote the Weyl nodes at 
		$\theta=$ 180\textdegree, 170\textdegree~and 160\textdegree,respectively.
		Dashed arrows show the motion of Weyl points with decreasing $\theta$. At $\theta=$180\textdegree, (collinear case) the two charge-2 Weyl nodes lie on the $C_{4z}$-protected $k_z$-axis. For $\theta \neq$180\textdegree~each charge-2 Weyl node splits into two charge-1 Weyl nodes which in turn move away from the $k_z$-axis. The integer above each Weyl node denotes the chiral charge.\label{wpevo}}
\end{figure}

Neutron scattering experiments have reported\cite{KRode13} a noncollinear ferrimagnetic magnetic structure in the DO$_{22}$ ferrimagnetic Mn$_3$Ga structure, where there is a significant in-plane magnetic moment, $\mu_x^{2b}$ = 1.19$\mu_B$ carried by the Mn$_I$ atoms [on the Mn-Ga (001) plane] leading to a 21$^{\circ}$ tilt of the Mn$_I$ moment from the crystallographic $c$ axis [see Fig.~\ref{wpevo}(a)].  
This noncollinear magnetic ordering spontaneously breaks both the $C_{4z}$ and $M_z$ symmetry operations while only preserving $P$. 
Consequently, the $C_{4z}$-protected double Weyl fermion on the $k_z$ axis for the case of collinear ferrimagntism splits into two charge-1 Weyl fermions
which shift away from the $k_z$ axis. 
\begin{table}[t]
	\caption {Comparison of the values of the magnetic moments for the Mn$_I$ atoms for the noncollinear case from theory and experiment. $\mu_z^{2b}$ ($\mu_x^{2b}$) denotes the $z$ ($x$)-component of the magnetic moment of Mn$_I$.}\label{Trot}
	\begin{ruledtabular}
		\begin{tabular}{ccccc}
			Method&$\mu_z^{2b}$($\mu_B$/Mn)&$\mu_x^{2b}$ ($\mu_B$/Mn)&$\theta$(\textdegree)\\[2pt]
			\hline
			rotated Wannier&-2.60&-0.94&160\\
			GGA+SOC&-2.65&-0.95&160\\
			experiment\cite{KRode13}&-3.07&1.19&159
		\end{tabular}	
	\end{ruledtabular}
\end{table}

In order to investigate this scenario, we have studied the evolution of the Weyl points upon rotation of all magnetic moments of the Mn$_I$ atoms at the Wyckoff positions 2b with respect to 
the crystallographic $z$ axis by the angle $\theta$,  ${\bm {\mu}}^{2b} = \mu^{2b} (-\sin\theta\hat x+\cos\theta\hat z)$, while fixing the direction of the Mn$_{II}$ magnetic moments, as shown in Fig.~\ref{wpevo}(a). Here,   $\theta=$ 180\textdegree~indicates the collinear (001) ferrimagnetism. 
Using the Wannier functions we find that at $\theta$ = 160\textdegree~
the magnitude of the calculated $x$-component of the magnetic moment of the Mn$_I$ atoms is 0.94$\mu_B$/Mn in good agreement with the corresponding experimental values of 1.19$\mu_B$. Table~\ref{Trot} summarizes the comparison of the values of the magnetic moments of the Mn$_I$ atom between theory and experiment. The magnetic moments from the rotated Wannier approach agree with DFT calculations well.
Fig.~\ref{wpevo}(b) shows the evolution of the Weyl nodes as $\theta$ changes from 180\textdegree~to 170\textdegree~ and finally to 160\textdegree. Initially, at  $\theta=$ 180\textdegree, the two charge-2 Weyl nodes lie on the $k_z$-axis. 
As $\theta$ decreases each charge-2 Weyl node splits into two charge-1 Weyl nodes which move away from the $k_z$-axis, leading to the emergence of four charge-1 Weyl fermions in the case of noncollinear ferrimagnetism. Our electronic structure calculations of the Fermi arcs on the (100) surface for $\theta$  = 160 \textdegree show that the noncollinear effect is small on the Fermi arcs in Fig.~\ref{WP}(b), at least for small angle.
 

\section{discussion}
In this section we discuss the effect of electron-electron interactions, U, on the equilibrium lattice constants, magnetic moments, and the topology of the band structure for the collinear magnetic structure. 
As was alluded earlier we employed U=2.6 eV for Mn$_I$ atoms and U=0 for the remaining atoms, which were found\cite{saha2017impact} 
to give a value for the $a$ lattice constant of 3.91 \AA~ in good agreement with the experimental value. However, as shown in Table~\ref{Tlat}, this is at the expense of a worse agreement for both the lattice parameter $c$ and the $z$-component of the magnetic moment of the Mn$^{2b}$ (Mn$_I$) atoms. 
As shown in Fig.~\ref{NL}(a) in the absence of SOC the presence of U shifts 
the position of the Weyl nodes (red dots) to higher energies and slightly towards the center of BZ. Moreover, even in the presence of SOC, the nodes are robust and remain gapless at about the same  energies. Therefore, the charge-2 Weyl nodes survive in the presence of electronic correlations. Moreover, since the effect of electronic correlations can not gap out  the charge-2 Weyl nodes, the charge-1 Weyl nodes splitting of the charge-2 Weyl nodes should be robust in the noncollinear magnetic structure as well.

 	
Due to the large shift of the Weyl nodes to higher energies induced by U, 
it will be challenging to observe the Fermi arcs above the Fermi level 
in Fig.~\ref{WP}(b) employing angle-resolved photoemission spectroscopy (ARPES). On the other hand, the time-resolved ARPES (trARPES), a fast-growing and powerful technique to observe conduction electron states up to hundreds meV above the Fermi level\cite{JASobota12, HSoifer19}, may be a suitable platform to observe the Fermi arcs above the Fermi level on the (100) surface in future experiments. Moreover, electron doping or alloying that preserves the $C_{4z}$ symmetry, e.g., Mn$_3$Ge in the cubic structure\cite{DZhang13}, can rise the chemical potential which may allow in turn the observation of 
these nontrivial surface states emerging from the Weyl nodes.

\section{CONCLUSION}
In summary, our {\it ab initio} electronic structure calculations have shown that in
the absence of SOC, nontrivial nodal lines emerge in collinear  ferrimagnetic tetragonal Mn$_3$Ga. The nodal lines are protected by both 
mirror reflection symmetry normal to the [110] direction, $M_{[110]}$,  and a four-fold rotational symmetry, $C_{4z}$. The presence of SOC gaps out the nodal lines except for the nodal line intersecting points which become $C_{4z}$-protected charge-2 Weyl nodes with quadratic dispersion in the $k_x$-$k_y$ plane. The noncollinear magnetism associated with the Mn$_I$ atoms splits 
the double Weyl nodes Fermions into two charge-1 Weyl nodes moving away from the $k_z$ axis, whose separation can be selectively tuned by the noncollinarity angle. 


\begin{acknowledgments}
The work at CSUN is supported by NSF-Partnership in Research and Education in Materials (PREM) Grant No. DMR-1828019. H.L. acknowledges the support by the Ministry of Science and Technology (MOST) in Taiwan under grant number MOST 109-2112-M-001-014-MY3.
\end{acknowledgments}


\begin{thebibliography}{10}
\bibitem{com1}
Our GGA+U+SOC electronic structure calculations reveal that the Weyl nodes are robust and remain gapless even in the presence of U.
\end{thebibliography}
\end{document}